\documentclass[10pt]{article}
\usepackage[cp1251]{inputenc}
\usepackage[dvips]{epsfig}   \usepackage[dvips]{changebar}
\usepackage[dvips]{graphicx}

\def\lsim{\
  \lower-1.2pt\vbox{\hbox{\rlap{$<$}\lower5pt\vbox{\hbox{$\sim$}}}}\ }
\def\gsim{\
  \lower-1.2pt\vbox{\hbox{\rlap{$>$}\lower5pt\vbox{\hbox{$\sim$}}}}\ }
   
  \usepackage{amsmath,enumerate, amsfonts, amssymb,amsthm}

\begin{document}
\title{Exact crystalline solution for a one-dimensional few-boson system with point interaction}
 \author{Maksim Tomchenko
\bigskip \\ {\small Bogolyubov Institute for Theoretical Physics} \\
 {\small 14b, Metrolohichna Str., Kyiv 03143, Ukraine} }
 \date{\empty}
 \maketitle
 \large
 \sloppy
 \textit{We study the exact solutions for a one-dimensional system of $N=2; 3$ spinless point bosons for zero boundary conditions.
 In this case, we are based on M. Gaudin's formulae obtained with the help of Bethe ansatz.
 We find the density profile $\rho(x)$ and the nodal structure of a wave function for a set of the lowest states
 of the system for different values of the coupling constant $\gamma\geq 0$.
The analysis shows that the ideal crystal corresponds to the quantum numbers (from Gaudin's equations) $n_{1}=\ldots =n_{N}=N$
 and to the coupling constant $\gamma\lsim 1$. We also find that the ground state of the system ($n_{1}=\ldots =n_{N}=1$)
 corresponds to a liquid for any $\gamma$  and any $N\gg 1$. In this case,
 the wave function of the ground state is nodeless, and the wave function of the ideal crystal has nodes.  } 

 \section{Introduction}

The one-dimensional (1D) crystals were observed in the form of
stripes in a two-dimensional (2D) system in many experiments (see,
e.g., \cite{keller1986,kashuba1993,vanderbildt1995} and references
therein). Recently, the stripe 1D Bose crystals were found in a
three-dimensional (3D) system placed in the field of a trap
\cite{ketterle2017,pfau2017,tanzi2019,pfau2019}. The 1D Bose
crystals are of great theoretical interest since a one-dimensional
system is much simpler than  two- and three-dimensional ones, which
facilitates the finding of solutions. On the whole, the analytic or
numerical search for an \textit{accurate} solution for a crystal is
a very complicated problem. All solutions we know for a crystal are
approximate.  In what follows, we have obtained (seemingly, for the
first time) the \textit{exact} solution for the ideal Bose crystal
of $N=2; 3$ point atoms in one dimension.

The solutions for 1D Bose crystals without the external field were
investigated in many works
\cite{keller1986,kashuba1993,vanderbildt1995,gross1958,nep,shevchenko1987,lozovik2005,shlyapa2015,andreev2017,boronat2017,boninsegni2019,mt2020}.
In particular, E.~Gross obtained a crystalline solution for a 1D
Bose system in the mean-field approximation \cite{gross1958}.
Yu.~Nepomnyashchii \cite{nep} considered the crystal-like 1D
solution for a 3D Bose system with a condensate of atoms with
momentum $\hbar \textbf{k}_{r}$ (where $\textbf{k}_{r}$ is a vector
of the reciprocal lattice of a 1D crystal). In the work by
S.~Shevchenko, a possibility of the appearance of a 1D superfluidity
along the line of dislocations in a 3D Bose crystal was studied
\cite{shevchenko1987}.  A.~Arkhipov et al. numerically found a
crystalline regime for the ground state (GS) of a 1D Bose system
with the dipole-dipole repulsive interaction \cite{lozovik2005}. The
crystalline 1D solutions for a 2D system of dipoles were obtained
within various methods in works
\cite{keller1986,kashuba1993,vanderbildt1995,shlyapa2015,andreev2017,boronat2017,boninsegni2019}.
The approximate 1D crystalline solution corresponding to a
fragmented condensate of atoms was found  for a $j$-dimensional
periodic system; $j=1; 2; 3$ \cite{mt2020} (the one-dimensional
nature of this solution can be seen from the fact that the axis $x$,
say, can always be directed along the vector $\textbf{k}$ describing
the crystal).

In the last decade, the 1D crystalline solutions were obtained for
3D \cite{pfau2017,pfau2019,ferlaino2019} and 1D
\cite{reimann2010,zollner2011,zollner2011b,chatterjee2013,lode2018,chatterjee2019}
systems of dipoles in a trap (see also the review on few-boson 1D
systems \cite{sowinski2019}). The GS of a 1D dipolar system
corresponds to a crystal in the case of the strong repulsion of
dipoles, which takes place at a large constant of dipole-dipole
interaction
\cite{reimann2010,zollner2011,zollner2011b,chatterjee2013,lode2018,chatterjee2019}
or at a small distance between dipoles
\cite{zollner2011,zollner2011b}. However, for a 1D Bose system with
the point interaction, GS corresponds to a gas (liquid) at any
parameters of the system. This was shown for an infinite system with
periodic boundary conditions (BCs) in the classical work by E. Lieb
and W. Liniger \cite{ll1963}. This property is true also for zero
BCs, since the GS energy of a 1D system of point bosons is the same
in the thermodynamic limit for  periodic and zero BCs
\cite{mt2015,syrwid2020}. In what follows, we will show that GS of a
\textit{finite} 1D system of point bosons with zero BCs also
corresponds always to a gas (liquid).

According to the well-known result by L. Landau, the infinite 1D
crystal must be destroyed by fluctuations \cite{land1937,land5}.
This conclusion is valid for the 1D crystalline ordering in a system
of any dimensionality: 1, 2 or 3. However, as early as Ya. Frenkel
noted that Landau's conclusion is not quite correct for real
(finite) crystals, because for a finite 1D system the transition
from the sum $\sum_{j=1}^{\infty}T/k_{j}^{2}$ to the integral
$(TL/2\pi)\int_{0}^{\infty}dk/k^{2}$ is not proper \cite{frenkel}
(this was also noted in \cite{land5}). In Nature, all systems are
finite. Therefore, the Landau's result significantly  reduces the
stability region  of 1D crystals, but does not forbid their
existence. In addition, the ground state of a 1D system of point
bosons corresponds to a liquid even at small $N, L$ (see the upper
curve in Fig. 5 below) for which Landau's forbidding clearly does
not work. Therefore, it is clear that the liquid character of GS of
a 1D system of point bosons is not related to Landau's forbidding.

In the present work, we will investigate the exact crystalline
solution and the nature of the lowest state for a 1D system of
spinless bosons with point interaction.

\section{Solution for a 1D Bose crystal}
It is impossible to find the exact solution for a 2D or 3D crystal,
at least at the present time. Some authors believe that certain
Monte Carlo methods provide a numerically exact solution. As far as
we see, in practice they give only an approximate solution.
Moreover, any  ``exact'' Monte Carlo method gives no analytic
solution for WF and observable quantities. However, in the case of
1D system, the exact crystalline solution can be found for a point
interatomic interaction. To this end, we use a theory based on the
Bethe ansatz \cite{ll1963,bethe,gaudin1971,gaudinm,takahashi1999}
(see also the recent reviews
\cite{syrwid2020,batchelor2014,guan2015}).

Consider the system of $N$ spinless point bosons not placed in an
external field, under zero BCs ($\Psi=0$ on the boundaries). The
Schr\"{o}dinger equation for such a system reads
\begin{equation}
 -\sum\limits_{j=1}^{N}\frac{\partial^{2}}{\partial x_{j}^2}\Psi + 2c\sum\limits_{l<j}
\delta(x_{l}-x_{j})\Psi=E\Psi.
     \label{god-0} \end{equation}
Here, $\hbar=2m=1$. Using the analysis in \cite{gaudin1971,gaudinm},
we find that, for $N=2,$ any pure state of such a system  is
described by WF
\begin{eqnarray}
 &&\Psi_{\{k \}}(x_{1},x_{2})|_{x_{1}\leq x_{2}}=4\left
 (1+\frac{c^{2}}{k_{1}^{2}-k_{2}^{2}}\right )\sin{k_{1}x_{1}}\sin{k_{2}x_{2}}+4\left
 (1+\frac{c^{2}}{k_{2}^{2}-k_{1}^{2}}\right
 )\sin{k_{2}x_{1}}\sin{k_{1}x_{2}} \nonumber \\ &&+ 4\left
 (\frac{c}{k_{1}-k_{2}}-\frac{c}{k_{1}+k_{2}}\right )\sin{k_{1}x_{1}}\cos{k_{2}x_{2}}
+4\left
 (\frac{c}{k_{2}-k_{1}}-\frac{c}{k_{1}+k_{2}}\right )\sin{k_{2}x_{1}}\cos{k_{1}x_{2}}.
      \label{god-1} \end{eqnarray}
If $x_{1}>x_{2}$, the solution for WF follows from (\ref{god-1}) by
the permutation of $x_{1}$ and $x_{2}$. For any $N\geq 2$, the
quantities $k_{1},\ldots,k_{N}$ satisfy the Gaudin's equations
\cite{gaudin1971,gaudinm}
\begin{eqnarray}
Lk_{j}=\pi n_{j}+\sum\limits_{l=1}^{N}\left
(\arctan{\frac{c}{k_{j}-k_{l}}} +
\arctan{\frac{c}{k_{j}+k_{l}}}\right )|_{l\neq j}, \ j=1,\ldots, N,
     \label{god-3} \end{eqnarray}
where $n_{j}$ are integers,  $n_{j}\geq 1$
\cite{gaudin1971,gaudinm,mtjpa2017}, $c\geq 0$ is the interatomic
interaction constant, and  $L$ is the size of the system. Different
states correspond to different collections of $\{n_{j}\}\equiv
(n_{1},\ldots,n_{N})$. In particular, $n_{j\leq N}= 1$ for GS. For
the excited states, $n_{j}\geq 1$ for all $j=1,\ldots,N$; in this
case, $n_{j}> 1$ at least for one $j$
\cite{mt2015,gaudin1971,gaudinm,mtsp2019}. We note that, in
(\ref{god-1})--(\ref{god-6}), $k_{1},\ldots,k_{N}$ denote the
quantities $|k_{1}|,\ldots,|k_{N}|,$ respectively.

For  $N=3$, WF of any pure state reads \cite{gaudin1971,gaudinm}
\begin{eqnarray}
 \Psi_{\{k \}}(x_{1},x_{2},x_{3})|_{x_{1}\leq x_{2}\leq x_{3}}=\sum\limits_{\varepsilon_{1},\varepsilon_{2},\varepsilon_{3}=\pm 1}
A(\varepsilon_{1},\varepsilon_{2},\varepsilon_{3})\sum\limits_{P}
a(k_{1},k_{2},k_{3})e^{i\varepsilon_{1}k_{1}x_{1}+i\varepsilon_{2}k_{2}x_{2}+i\varepsilon_{3}k_{3}x_{3}},
      \label{god-4} \end{eqnarray}
where $\sum_{P}$ is the sum over all possible permutations of the
numbers
$\varepsilon_{1}k_{1},\varepsilon_{2}k_{2},\varepsilon_{3}k_{3}$,
\begin{eqnarray}
a(k_{1},k_{2},k_{3})=\left
(1+\frac{ic}{\varepsilon_{1}k_{1}-\varepsilon_{2}k_{2}}\right )\left
(1+\frac{ic}{\varepsilon_{2}k_{2}-\varepsilon_{3}k_{3}}\right )\left
(1+\frac{ic}{\varepsilon_{1}k_{1}-\varepsilon_{3}k_{3}}\right ),
     \label{god-5} \end{eqnarray}
\begin{eqnarray}
A(\varepsilon_{1},\varepsilon_{2},\varepsilon_{3})=\varepsilon_{1}\varepsilon_{2}\varepsilon_{3}\left
(1-\frac{ic}{\varepsilon_{1}k_{1}+\varepsilon_{2}k_{2}}\right )\left
(1-\frac{ic}{\varepsilon_{2}k_{2}+\varepsilon_{3}k_{3}}\right )\left
(1-\frac{ic}{\varepsilon_{1}k_{1}+\varepsilon_{3}k_{3}}\right ).
     \label{god-6} \end{eqnarray}
WFs (\ref{god-1}) and (\ref{god-4}) are not normalized. Note that WF
(\ref{god-1}) is real, and WF (\ref{god-4}) is purely imaginary. In
order to obtain $\Psi_{\{k \}}(x_{1},x_{2},x_{3})$ for the
configuration $x_{3}\leq x_{1}\leq x_{2}$ (for example), one needs
to make in (\ref{god-4}) changes $x_{1}\rightarrow x_{3}$,
$x_{2}\rightarrow x_{1}$, $x_{3}\rightarrow x_{2}$.

We remark that, in the analysis below, the values of $k_{j}$ for
different $c, N, L, n_{j}$ are found by numerically solving Eqs.
(\ref{god-3}) by the Newton method.

The analysis of functions (\ref{god-1}) and (\ref{god-4}) for $N=2;
3$ indicates that the ground-state WF  is nodeless for any coupling
constant $\gamma\equiv c/g\geq 0$ (but $\gamma< \infty$), where $g=
N/L$ is the concentration (we follow the terminology \cite{gilbert},
according to which the nodal points, nodal lines, and nodal surfaces
of any dimensionality are called nodes). For the low-lying states
with the given collection of numbers $n_{j}$ (we studied all states
with $1\leq n_{1},\ldots,n_{N}\leq N+1$), the evolution of the nodal
structure of WF occurs in the following way, as $\gamma$ increases.
For $\gamma= 0,$ the nodal surfaces divide the space $0\leq
x_{1},\ldots,x_{N}\leq L$ into $l$ segments, where the value of $l$
depends on the collection $\{n_{j}\}$. At $0< \gamma < \infty$ for
some collections
 $\{n_{j}\}$, several of
those segments are united, whereas other segments can separate
themselves into parts. In this case, for each collection
$\{n_{j}\},$ the number of segments $l$ is the same for all $0<
\gamma < \infty$. As $\gamma$ increases, only the shape of segments
slightly changes (see Figs. 1, 3). Presumably, these properties hold
true for the high-lying states as well.

\begin{figure}
\includegraphics[width=0.47\textwidth]{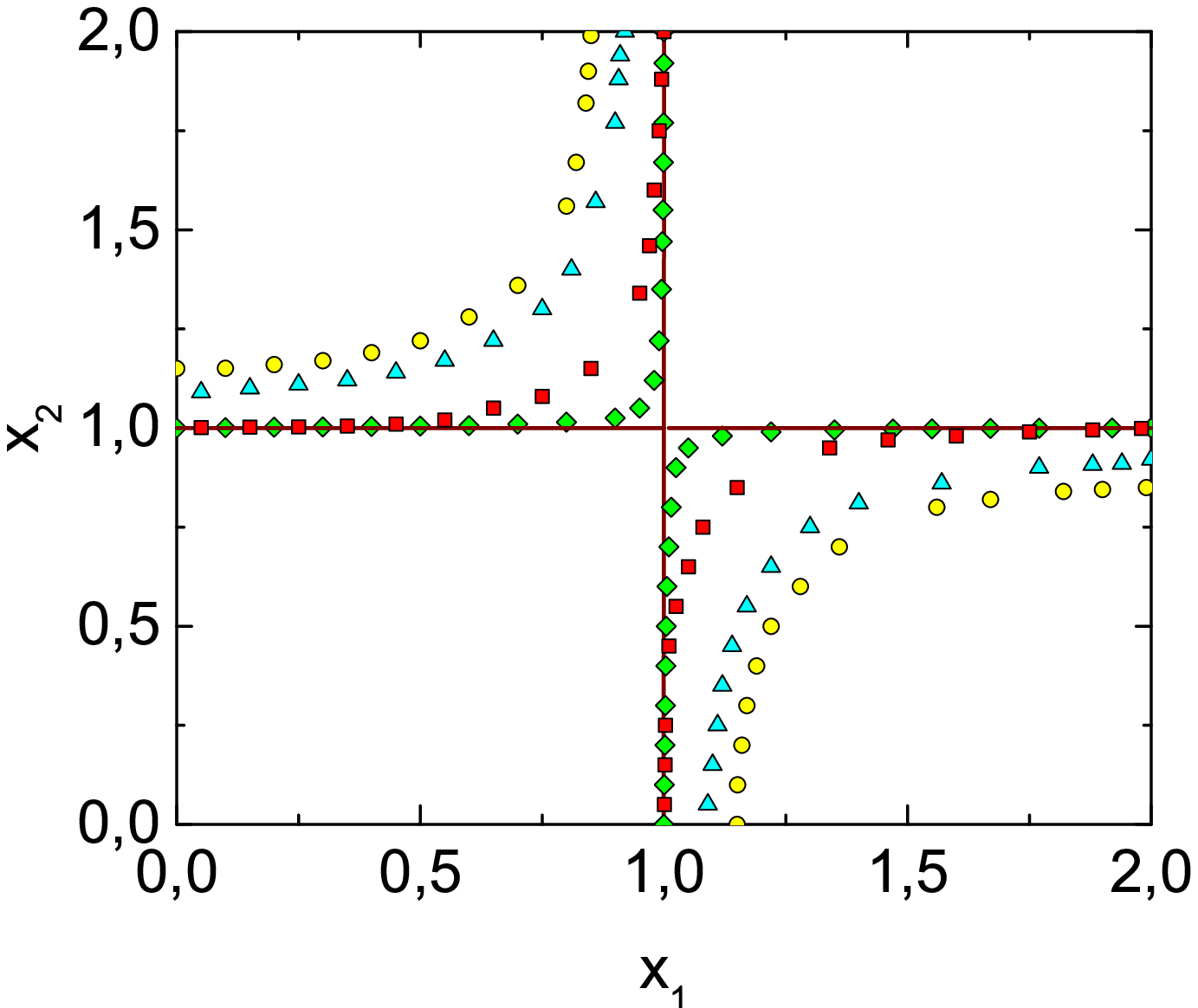}
\hfill
\includegraphics[width=0.47\textwidth]{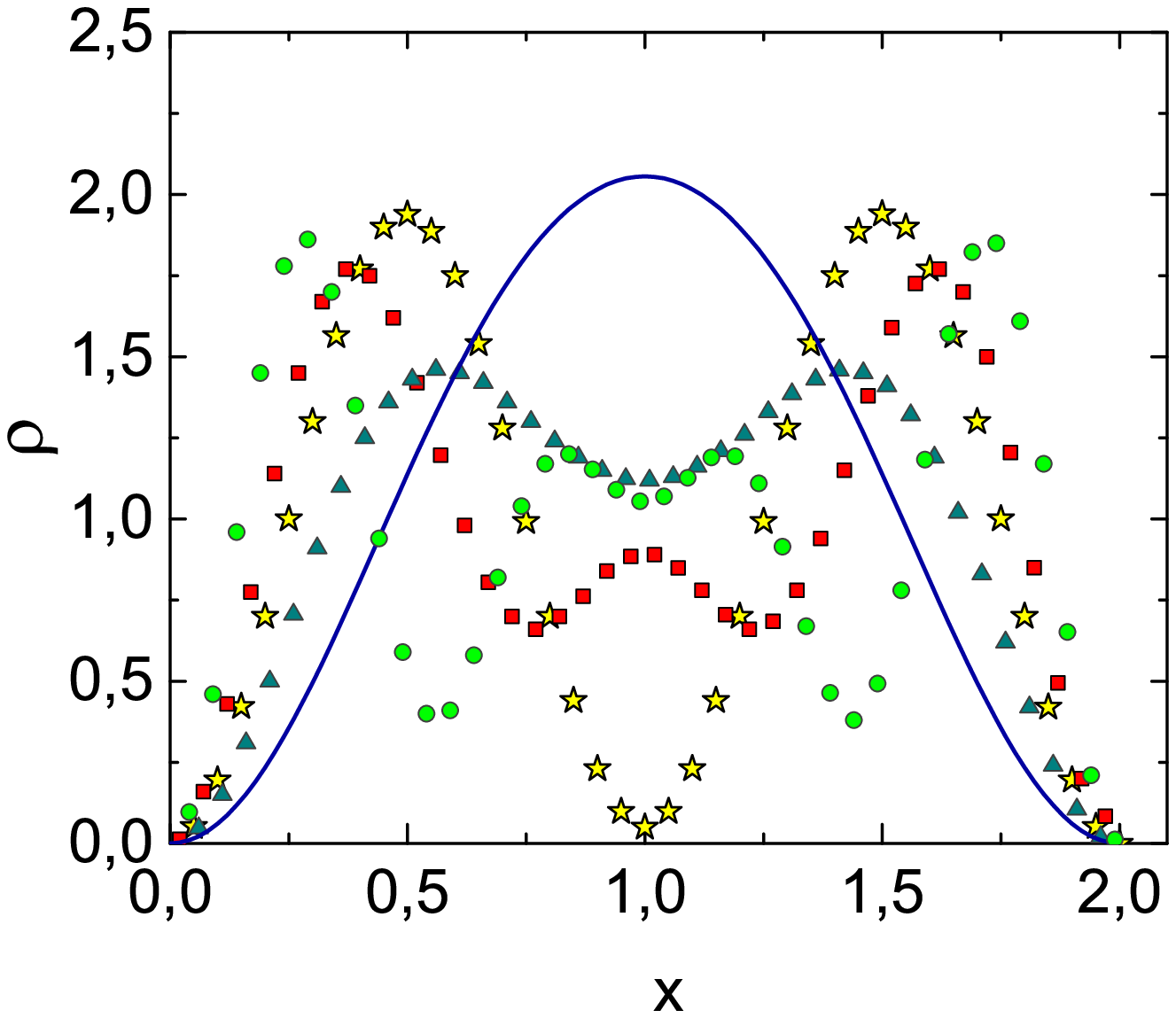}
\\
\parbox[t]{0.47\textwidth}{
\caption{[Color online] Nodal lines ($\Psi(x_{1},x_{2})=0$) for a
system of $N=L=2$ point bosons under the zero BCs for the state with
$n_{1}=n_{2}=2$: at $\gamma=0$ (solid lines), $\gamma=0.1$ (rhombs),
$\gamma=1$ (squares), $\gamma=10$ (triangles), and $\gamma=100$
(circles).
 \label{fig1}} } \hfill
\parbox[t]{0.47\textwidth}{
\caption{[Color online] Density profile   $\rho(x)$ of a system of
point bosons under the zero BCs at $N=L=2$, $\gamma=1$, for the
following states: $n_{1}=n_{2}=1$ (solid line); $n_{1}=1, n_{2}=2$
(triangles); $n_{1}=n_{2}=2$ (stars); $n_{1}=2, n_{2}=3$ (squares);
and $n_{1}=3, n_{2}=4$ (circles). \label{fig2}} }
\end{figure}

Thus, the nodal structure of WF is  the same for all $0< \gamma <
\infty$. As $\gamma $ increases, some segments merge, while others
divide themselves into parts. In this case, the maxima of
$|\Psi|^{2}$  smoothly shift. The transition from $\gamma \gg 1$ to
$\gamma = \infty$ leads to the addition of the nodal lines
$x_{j}=x_{l}$ (here, $j\neq l$ and $j,l=1,\ldots,N$). In general,
for each state $(n_{1},\ldots,n_{N})$ the distribution of atoms in a
rough approximation is approximately the same for all $0\leq \gamma
\leq \infty$. This means that the structure of any solution,
including the crystalline one, for a given  $\gamma > 0$ can be very
approximately understood by examining the solution for free bosons
($\gamma = 0$) with the same $n_{1},\ldots,n_{N}$ (we stress that we
keep in mind the structure of the solution in a crude approximation:
the nodal structure of WF and approximate distribution of atoms; to
understand the details of the solution, one needs, of course, to
find the solution for a given $\gamma
> 0$, as one can see in Fig. 4). We suppose that this property is true
for any $N\geq 2$ and any dimension of the space. However, for a
non-point
interaction this property is  not valid, at least for the 1D systems
(e.g., GS of a 1D system can correspond to either a crystal, or
liquid, depending on the coupling constant
\cite{reimann2010,zollner2011,zollner2011b,chatterjee2013,lode2018,chatterjee2019}).

Interestingly, the ideal crystal (the state of the $N$-particle
system for which the density profile $\rho(x)=N\int dx_{2}\ldots
dx_{N}|\Psi(x,x_{2},\ldots,x_{N})|^{2}$ contains $N$ identical
equidistant peaks) corresponds to the quantum numbers
$n_{1}=n_{2}=\ldots =n_{N}=N$ and to values $\gamma \lsim 1$ (see
Figs. 2 and 4). In Fig. 2 we present the  density profiles $\rho(x)$
at $N= 2$ for the ground state ($n_{1}=n_{2}=1$) and states close to
the crystal one.  Fig. 4 shows the evolution of the  density profile
for  $N= 3$, as $\gamma$ increases. In Figs. 2 and 4, $\Psi$ is
normalized to $1$. For $N> 3$, the ideal crystal should also
correspond to the quantum numbers $n_{j\leq N}=N$ since, for free
bosons, the crystal corresponds namely to such $n_{j}$ (as can be
shown), and with an increase in $\gamma $
the nodal structure  almost does not change (see above). The formula
$n_{j\leq N}=N$ is also related to results by A. Syrwid and K. Sacha
\cite{syrwid2017}. In this article, the averaged density
$\tilde{\rho}(x)$ is found for the state $n_{1}=\ldots =n_{N}=j$ at
$j=2; 3; 4$ and $N=6; 7$ (see also review \cite{syrwid2020}). As far
as we understand, $\tilde{\rho}(x)$ from \cite{syrwid2017} is
equivalent to $\rho_{N}(x_{N})$ from \cite{syrwid2015}: it is the
probability density of finding the $N$-th particle at the point
$x_{N}$ provided that the coordinates of $N-1$ remaining particles
have been measured, and the averaging over many measurements has
been carried out. Perhaps, under the zero BCs $\tilde{\rho}(x)$
coincides with $\rho(x)$, if $\tilde{\rho}(x)$ is obtained by
averaging over an infinite number of measurements. Each solution in
\cite{syrwid2017} contains $j$ immovable domains associated by the
authors with $j-1$ solitons. In our opinion, the analysis in
\cite{syrwid2017} is not sufficient for the proof of solitonic
properties of such solutions. In particular, the classical solitonic
analogs of such solutions have not been found.  Rather, these are
domain solutions.  The solution with $j=N$ corresponding to the
ideal crystal was not considered in \cite{syrwid2017}. However, the
results in \cite{syrwid2017} indicate that for $j=N$ there should be
$N$ domains. This agrees with the above results, though we have
found the ordinary density $\rho(x)$ instead of $\tilde{\rho}(x)$
(under periodic BCs, the density is constant: $\rho(x)=const$;
therefore, to determine the structure of a solution, one needs to
find more complicated functions instead of $\rho(x),$ for example,
the binary distribution function $g_{2}(x_{1},x_{2})$ or
$\tilde{\rho}(x)$; however, under the zero BCs, we have $\rho(x)\neq
const$ and  it is usually sufficient to find only $\rho(x)$ to
understand the structure of a solution).

Based on available information, we suppose that the \textit{ideal}
crystal corresponds solely to the quantum numbers
$n_{1}=n_{2}=\ldots =n_{N}=N$ (i.e., any other collection
$\{n_{j}\}$ does not correspond to the ideal crystal, for any
$\gamma$).

By the node theorem  \cite{gilbert}, if the states $\Psi_{j}$ are
associated with the numbers $j=1,2,\ldots,\infty$ in the order of
increasing energies $E_{j}$, then the nodes of the function
$\Psi_{j}$ divide the phase space $x_{1},\ldots,x_{N}$ into $l\leq
j$ segments. In this case, the lowest state must be non-degenerate,
and the higher states can be degenerate \cite{gilbert}. The theorem
was proved for one particle in the 2D space, but the proof can be
easily generalized to the case of any $N>1$ and any dimension of the
space. Our results for $N=2; 3$ are in agreement with this theorem.
Additionally, we have established for the lowest states of a 1D
system of $N=2; 3$ point bosons that, for any  collection
$\{n_{j}\},$ the number of segments ($l$) is independent of $\gamma$
(if $0<\gamma<\infty$), and the shape of segments depends weakly.

\begin{figure}
\includegraphics[width=0.47\textwidth]{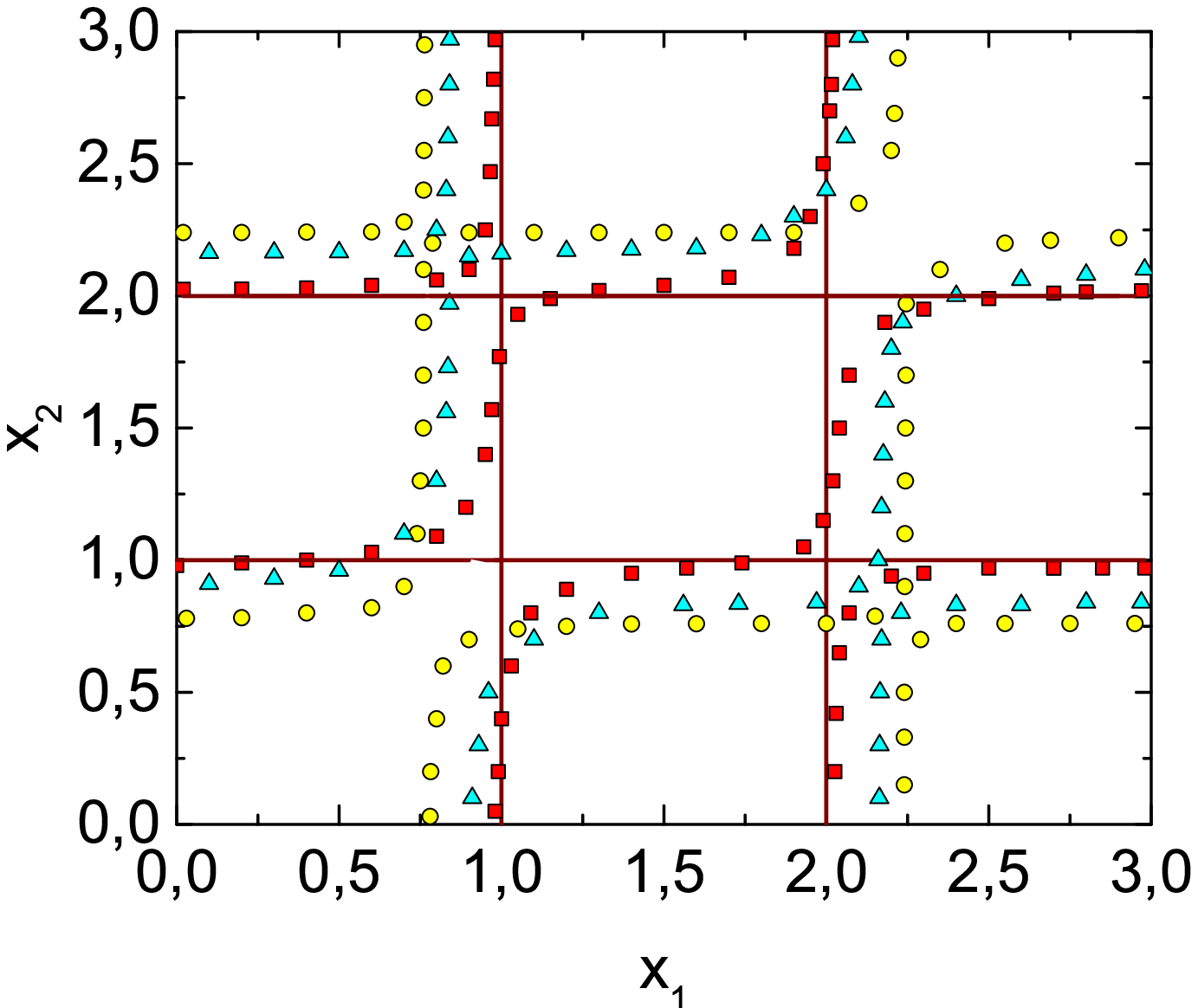}
\hfill
\includegraphics[width=0.47\textwidth]{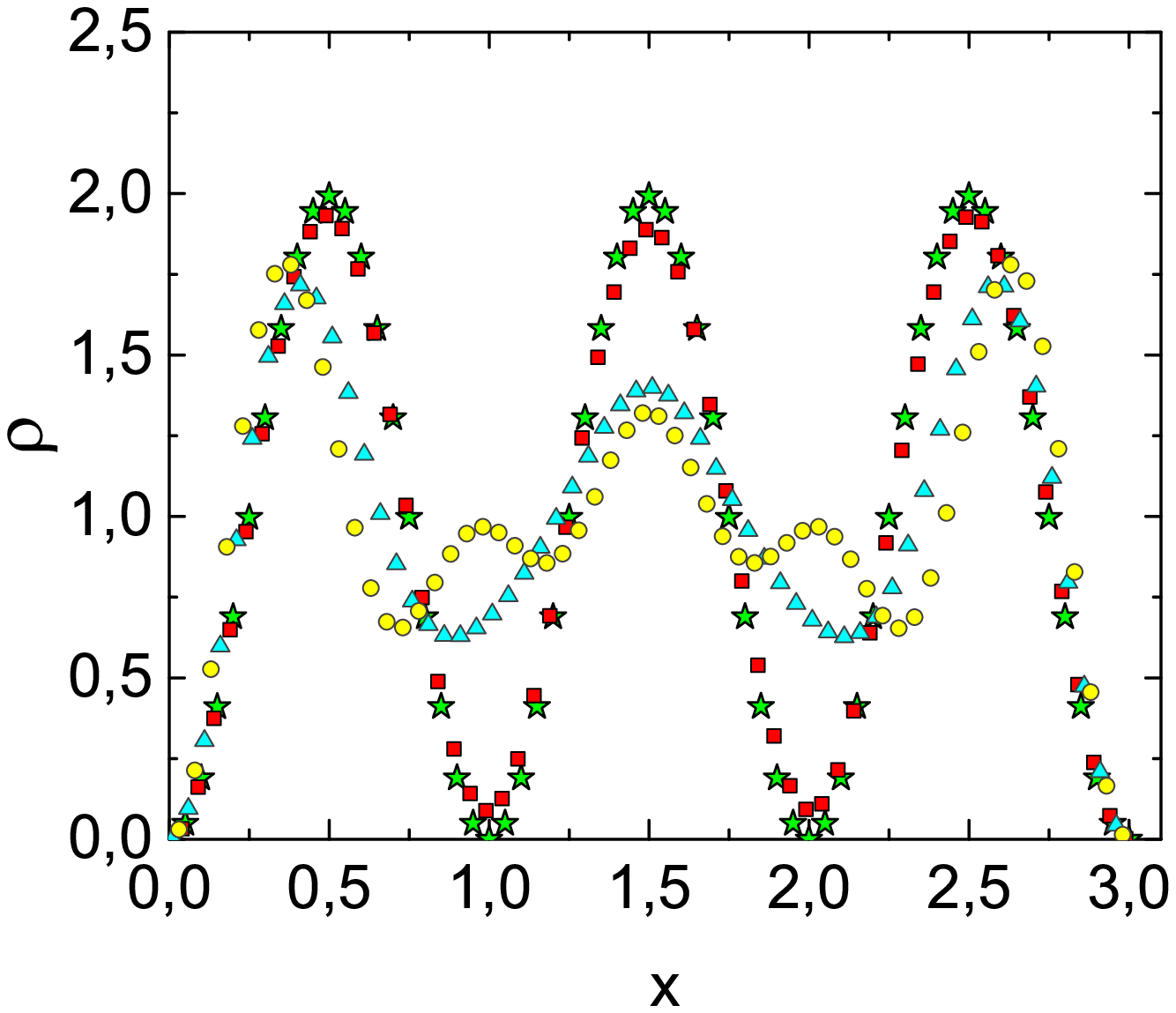}
\\
\parbox[t]{0.47\textwidth}{
\caption{[Color online] Nodal lines
($\Psi(x_{1},x_{2},x_{3})|_{x_{3}=L/2}=0$) for a system of $N=L=3$
point bosons under the zero BCs for the state with
$n_{1}=n_{2}=n_{3}=3$: for $\gamma=0$ (solid lines), $\gamma=1$
(squares), $\gamma=10$ (triangles), and $\gamma=100$ (circles).
Here, we show the cross-section of a 3D nodal network by the plane
$x_{3}=L/2$.
 \label{fig3}} } \hfill
\parbox[t]{0.47\textwidth}{
\caption{[Color online] Density profile $\rho(x)$ of a system of
point bosons under the zero BCs  for the state with
$n_{1}=n_{2}=n_{3}=3$, $N=L=3$: for $\gamma=0$ (stars), $\gamma=1$
(squares), $\gamma=10$ (triangles), and $\gamma=100$ (circles).
\label{fig4}} }
\end{figure}

For illustration, we remark that the nodes can ``draw''  whimsical
patterns in the profile of the probability density
$|\psi_{j}(x_{1},\ldots,x_{N}|^{2}$, see Figs. 5 and 6 in
\cite{lode2020uzory}.

From a physical standpoint, the properties described above mean the
following. Each $n_{j}>1$ corresponds to a quasiparticle with a
quasimomentum $p_{j}=\hbar \pi (n_{j}-1)/L$ \cite{mtsp2019}. At the
weak and strong coupling, we have, respectively, Bogolyubov--Feynman
quasiparticles \cite{bog1947,bz,fey1972} and Girardeau--Lieb
particle excitations \cite{girardeau1960,lieb1963}. Therefore, each
state $\{n_{j}\}$ corresponds to some collection of quasiparticles
$\{p_{j}\}$. According to the above analysis, for each state with a
collection of quasiparticles $\{p_{j}\},$ the distribution of atoms
(maxima of $|\Psi|^{2}$) is approximately the same for all $0\leq
\gamma \leq\infty$. The ideal crystal corresponds to $n_{1}=\ldots
=n_{N}=N$, i.e., to $N$ identical quasiparticles with the
quasimomentum $p_{j}=\hbar \pi (N-1)/L\equiv p_{c}$. In other words,
the \textit{crystal is formed by the condensate of $N$
quasiparticles} with the quasimomentum $p_{c}$. It is the lowest
(ground) state of the crystal. The close states for which some
$n_{j}$ are slightly different from $N$ correspond presumably to a
crystal with excitations or defects.

It is worth noting that the quasiparticles in a 1D system of point
bosons are usually divided into ``holes''  and  ``particles'',
following the work by E. Lieb \cite{lieb1963}. However, it was
recently shown  that, at weak coupling, the hole is a set of
interacting phonons (``particles'') with the lowest momentum
\cite{holes2020}. This does not contradict the solitonic properties
of holes (these properties have been found at weak coupling for the
holes with a large momentum $p\gg\hbar \pi /L$
\cite{syrwid2015,ishikawa1980,sato2012,sato2016,brand2018}).

Thus, under the zero BCs, there exist the following states with a
condensate of quasiparticles: The hole state with a quasimomentum
$\hbar \pi l/L$ (where $l\sim N$) is a condensate of $l$ phonons
with a quasimomentum $\hbar \pi/L$. The $j$-domain state
\cite{syrwid2017} is a condensate of $N$ phonons with a
quasimomentum $\hbar (j-1)\pi/L$.  The above-obtained solution for
the ideal crystal corresponds to a condensate of $N$ short-wave
phonons with a quasimomentum $\hbar (N-1)\pi/L$. Such states have
not been found in experiments, to our knowledge. Most probably, all
(or almost all) of them are unstable without the support of an
external field with a similar domain structure. It is natural to
expect that each of the mentioned above states with a condensate of
phonons with a quasimomentum $\hbar p\pi/L$ ($p=1; j-1; N-1$) is
also characterized by a condensate of atoms with the same
quasimomentum.

\section{The lowest state of a Bose crystal: basic models}
Let us concisely consider how the GS of a Bose crystal is described
in the literature. Several ans\"{a}tze for the ground-state WF of a
crystal were investigated (see reviews
\cite{guyer,galli2008,chan2013}). The historically first and most
widely used is a localized-Jastrow ansatz
\cite{bernardes1960,saunders1962,nosanow1962,brueckner1965,nosanow1966,levesque1968}
 \begin{equation}
   \Psi^{c}_{0} = C e^{S_{0}}\sum\limits_{P_{c}} e^{\sum_{j=1}^{N}\varphi(\textbf{r}_{j}-\textbf{R}_{j})},
 \label{god-7}    \end{equation}
where $N$ is the number of atoms in the system, $\textbf{r}_j$ and
$\textbf{R}_j$ are, respectively, the coordinates of atoms and
lattice sites, $\varphi(\textbf{r})=-\alpha^{2} \textbf{r}^{2}/2$,
and $P_{c}$ denotes all possible permutations of the coordinates
$\textbf{r}_{j}$ [e.g., for $N=2$: $\sum_{P_{c}}
\exp{\{\sum_{j=1}^{N}\varphi(\textbf{r}_{j}-\textbf{R}_{j})\}}=e^{\varphi(\textbf{r}_{1}-\textbf{R}_{1})+\varphi(\textbf{r}_{2}-\textbf{R}_{2})}
+e^{\varphi(\textbf{r}_{1}-\textbf{R}_{2})+\varphi(\textbf{r}_{2}-\textbf{R}_{1})}$].
Here, $\textbf{R}_j$ are fixed and are the same for all possible
configurations $\{\textbf{r}_{j}\}$.

A wave-Jastrow ansatz reads \cite{woo1976,ceperley1978}
 \begin{equation}
 \Psi^{c}_{0} = C e^{S_{0}-\sum\limits_{j=1}^{N} \theta(\textbf{r}_{j})},
 \label{1-11}    \end{equation}
where the function $\theta(\textbf{r})$  is periodic with the
periods of a crystal. In ans\"{a}tze (\ref{god-7}) and (\ref{1-11}),
the function $S_{0}$ is usually written in the Bijl--Jastrow
approximation $S_{0} = \sum_{l <
j}S_{2}(\textbf{r}_{l}-\textbf{r}_{j})$ \cite{bijl,jastrow1955}.
Ansatz (\ref{1-11}) is a wave type solution and corresponds to a
condensate of atoms with WF $\Psi_{c}(\textbf{r})\simeq
e^{-\theta(\textbf{r})}$. In the case of the localized ansatz
(\ref{god-7}), a condensate of atoms is absent
\cite{penronz1956,leggett2006,prokofev2007}.

Later, a ``shadow'' ansatz was proposed
\cite{vitiello1988,reatto1998,reatto2009}:
 \begin{equation}
\Psi_{0}(r)=\int e^{-\Xi(r,s)}ds, \quad
r\equiv\textbf{r}_{1},\ldots,\textbf{r}_{N}.
 \label{1-12}    \end{equation}
Here,  $s\equiv\textbf{s}_{1},\ldots,\textbf{s}_{N}$  are ``shadow''
variables, and
$\Xi(r,s)=\sum_{j_{1}<j_{2}}u_{r}(|\textbf{r}_{j_{1}}-\textbf{r}_{j_{2}}|)+
\sum_{k}u_{sr}(|\textbf{r}_{k}-\textbf{s}_{k}|)+\sum_{j_{3}<j_{4}}u_{s}(|\textbf{s}_{j_{3}}-\textbf{s}_{j_{4}}|)$.
If $u_{r}, u_{sr}, u_{s}$ are represented in the form of appropriate
Fourier series, then WF (\ref{1-12}) is translationally invariant.

Finally, a translationally invariant ansatz is given by
\cite{mcmillan1965,chester1970,reatto1995}
\begin{equation}
\Psi_{0} =  C\cdot e^{S_{0}}, \label{1-13}    \end{equation}
\begin{eqnarray}
   S_{0} &=&  \sum\limits_{j_{1}<j_{2}}S_{2}(\textbf{r}_{j_{1}}-\textbf{r}_{j_{2}})+
\sum\limits_{j_{1}<j_{2}<j_{3}}S_{3}(\textbf{r}_{j_{1}}-\textbf{r}_{j_{2}},\textbf{r}_{j_{2}}-\textbf{r}_{j_{3}})+\ldots
\nonumber \\ &+&\sum\limits_{j_{1}<j_{2}<\ldots <
j_{N}}S_{N}(\textbf{r}_{j_{1}}-\textbf{r}_{j_{2}},\textbf{r}_{j_{2}}-\textbf{r}_{j_{3}},\ldots
, \textbf{r}_{j_{N-1}}-\textbf{r}_{j_{N}}).
 \label{1-14}    \end{eqnarray}
Function (\ref{1-12}) is a partial case of WF (\ref{1-13}),
(\ref{1-14}). The latter sets the general form of the exact
(isotropic or not) translationally invariant WF of the ground state
of the Bose system. In other words, the gaseous, crystalline, and
liquid GSs of a periodic Bose system are described by  WF
(\ref{1-13}), (\ref{1-14}). In practice,  it is usually restricted
to the simplest Bijl--Jastrow approximation; sometimes, a
three-particle correction $S_{3}$ is additionally taken into
account.

All ans\"{a}tze (\ref{god-7})--(\ref{1-14}) are applicable to 1D, 2D
and 3D crystals. Ans\"{a}tze (\ref{god-7}), (\ref{1-11}),
(\ref{1-12}) are approximate, and ansatz (\ref{1-13}), (\ref{1-14})
is exact. Ans\"{a}tze (\ref{god-7}), (\ref{1-11}) can be treated as
approximate solutions for a crystal with zero BCs. Ans\"{a}tze
(\ref{1-12}), (\ref{1-13}), (\ref{1-14}) are used under periodic
BCs. All  available in the literature calculations (analytical and
numerical) with ans\"{a}tze (\ref{god-7})--(\ref{1-14}) are
approximate. Therefore, it is still hard to make a reliable choice
between ansatz (\ref{god-7}) and ansatz (\ref{1-11}), say.

Ans\"{a}tze (\ref{god-7})--(\ref{1-14}) describe crystalline states
corresponding to the genuine (nodeless) GS of the Bose system. In 2D
and 3D cases, a nodeless crystalline solution is seemingly possible
for any reasonable interatomic potential: non-point or point-like.
However, for a 1D system the nodeless solutions
(\ref{god-7})--(\ref{1-14}) are possible only for a non-point
potential. Indeed, we have found above (Sect. 2) for the 1D system
of spinless point bosons that the WF of the ideal crystal
indispensably has nodes. The structure of the WFs of the crystal for
$N=2; 3$ shows that, in the thermodynamic limit ($N,
L\rightarrow\infty$, $N/L=const$), the exact WF of the ideal 1D
crystal with the zero BCs is, seemingly, quite close to the function
\begin{eqnarray}
 \Psi^{c}_{0}&=&C e^{S_{0}+S_{c}} \prod\limits_{j=1}^{N}
 \sin{(k_{c}x_{j})},
  \label{c-z}    \end{eqnarray}
where $k_{c}=N\pi/L=\pi/a$, $a$ is the period of the lattice, and
$S_{0}$ is equal to the sum of a two-particle term
$\sum_{p<l}S_{2}(x_{p}-x_{l})$  and the many-particle correlation
corrections. The function $S_{c}(x_{1}, \ldots, x_{N})$ is unknown
and can be found from the Schr\"{o}dinger equation. The product of
sines  in (\ref{c-z}) directly  sets the crystal lattice.

One can comprehend the structure (\ref{c-z}) from a qualitative
reasoning. Consider a 1D system of $N$ free bosons under the zero
BCs. In this case,  GS is described by the nodeless $\Psi_{0}=C
\prod_{j=1}^{N}
 \{\sin{(\pi x_{j}/L)}\}$, and the crystal is set by the multi-node WF (\ref{c-z}) with $S_{0}=S_{c}=0$.
It is an ideal crystal with one atom per cell. It was shown above
that if we ``switch on'' the point interaction and increase its
strength, then the number and location of cells do not virtually
change. Only the probability density distribution for atoms in the
cell changes. It holds for the interatomic interaction of zero range
and, apparently, small nonzero one. Therefore, we can crudely see
the general structure of WF of the crystal in the case of a strong
coupling by considering the corresponding WF for zero coupling.
Namely, the crystal is formed by a condensate of $N$ quasiparticles
with a quasimomentum $p_{c}=\hbar \pi (N-1)/L$. At zero coupling,
this corresponds to $N$ free bosons with a quasimomentum $p_{c}$.
We suppose that formula (\ref{c-z}) qualitatively correctly
describes the structure of GS of a macroscopic  finite crystal
(everywhere, except for a narrow near-boundary region, where the
size of cells can be different), if the interaction has a small
radius ($\ll\bar{d}$; $\bar{d}$ being the average interatomic
distance) and is repulsive. In this case, the factor $e^{S_{0}}$
describes the interatomic correlations, and the factor $e^{S_{c}}$
corresponds to the influence of the interatomic interaction on the
distribution of atoms in the cell.

We remark that, at the interatomic interaction with nonzero range,
the ``replacement'' of the solution for a strong coupling by the
corresponding one for zero coupling may lead to an error. Since for
such a 1D system with strong coupling, the crystal may correspond to
a nodeless $\Psi^{c}_{0}$ (the genuine GS of the system)
\cite{lozovik2005,reimann2010,zollner2011,zollner2011b,chatterjee2013,lode2018,chatterjee2019},
whereas the crystal-like solution $\Psi^{c}_{0}=C
\prod\limits_{j=1}^{N} \{\sin{(\pi x_{j}/a)}\}$ obtained for zero
coupling has many nodes.

It is currently accepted in the literature that the lowest states of
2D and 3D Bose crystals  are described by wave functions without
nodes. However, we do not exclude the existence in Nature of Bose
crystals whose lowest state corresponds to a wave function with many
nodes (which is similar to (\ref{c-z}), but is two-dimensional or
three-dimensional). However, there is yet no reliable evidence
(theoretical or experimental) of the existence of such crystals. In
particular, symmetry analysis is not enough to ascertain whether the
GS of a Bose system is a crystal or a liquid \cite{mt2022sym}.

\section{Discussion}
It is worth noting that the field of a trap can strongly change the
structure of GS, since the trap field decreases the symmetry of the
Hamiltonian $\hat{H}$. In particular, when switching on the trap
field, $\hat{H}$ ceases to commute with the operator of the total
momentum. In this case, the translational symmetry disappears. If
the trap field is very strong and corresponds to an ideal crystal,
then it is clear that the genuine GS of a system corresponds most
likely to the same crystal. On the whole, we may expect that GS
corresponds to the most symmetric WF from the set of eigenfunctions
of the boundary-value problem \cite{elliott}. Therefore, a decrease
in the symmetry of the Hamiltonian should lead to a decrease in the
symmetry of GS.

\subsection{Structure of the ground state of the 1D system of point bosons}

In Section 2, we considered the exact solutions for systems of $N=2;
3$ spinless point bosons. The results indicate that the ideal
crystal corresponds to WF with a lot of nodes, whereas the nodeless
GS is characterized by the most smooth density profile $\rho(x)$
and, therefore, corresponds to a liquid (gas). At $N\gg 1$, GS of
such a system also corresponds to a liquid (gas) for any $\gamma
\geq 0$. This is seen from that the GS energy  $E_{0}$
\cite{ll1963,mt2015,gaudin1971,bulatov1988} for $\gamma \lsim 1$ is
close to Bogolyubov's one \cite{bog1947,bz}, and the curves
$E_{0}(g)|_{N, \gamma=const}\sim g^{2}$ and $E_{0}(\gamma)|_{N,g=
N/L=const}$ \cite{ll1963} for $0< \gamma < \infty$ contain no
features, which indicate a possible transition to the crystal
regime. Finally, at $\gamma = \infty$, the system is a gas of
impenetrable bosons \cite{girardeau1960}.

\begin{figure*}
\includegraphics[width=.6\textwidth]{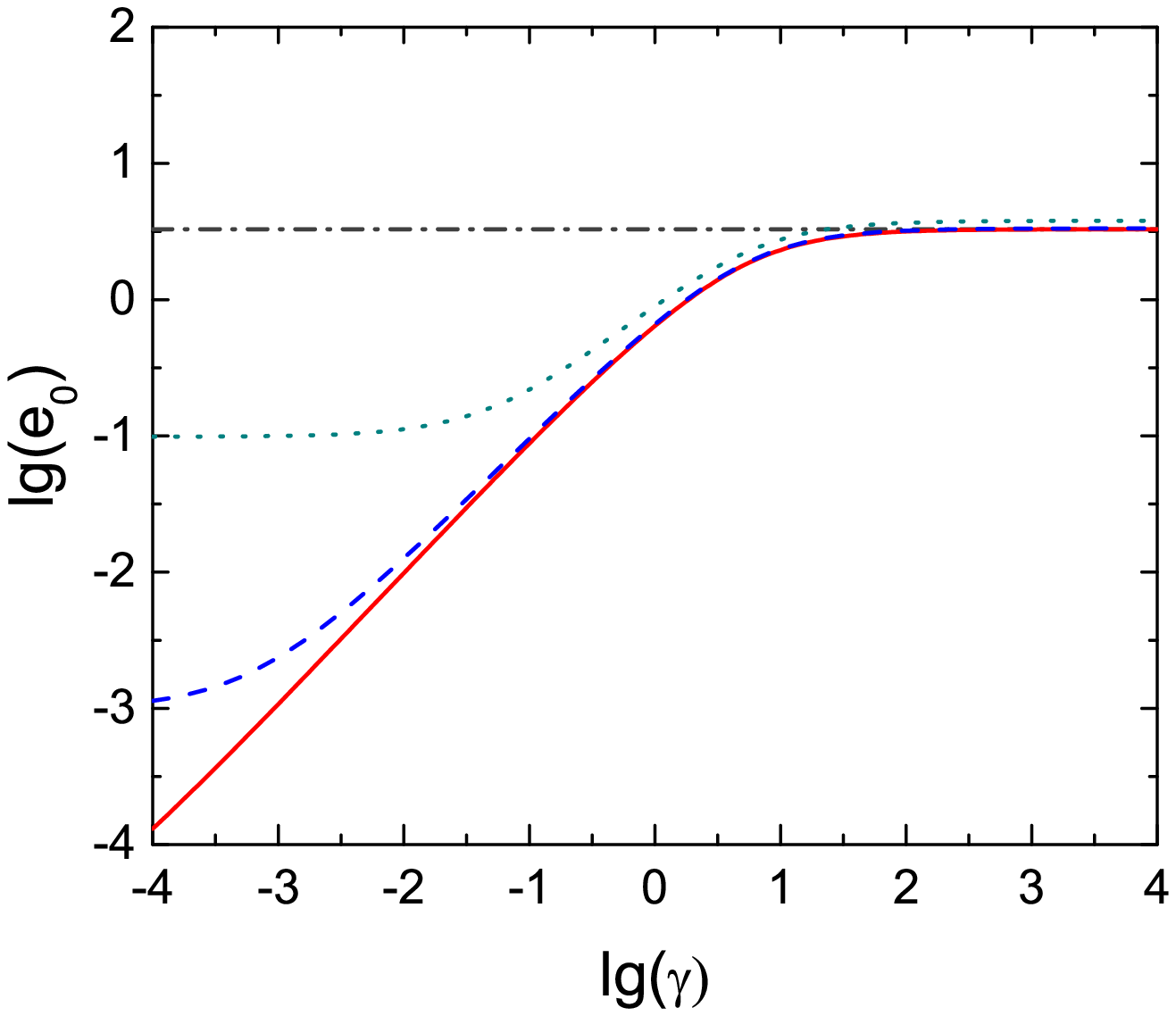}
\caption{[Color online] Function $e_{0}(\gamma)$ (\ref{e0ll}) for a
1D system of spinless point bosons with the zero BCs for $N=10$
(dotted line), $N=100$ (dashed line), and $N=1000$ (solid line).
Dash-dotted line marks the limiting value $e_{0}(\gamma\rightarrow
\infty, N=1000)=\left (1-\frac{1}{N^{2}} \right )\frac{\pi^{2}}{3}$
obtained by M. Girardeau for periodic BCs \cite{girardeau1960}.
Here, $\lg\equiv\log_{10}$. }
          \label{fig:5}                  \end{figure*}

The results listed in the previous paragraph are obtained mainly for
periodic BCs and in  the thermodynamic limit ($N, L\rightarrow
\infty$, $N/L=const$). We now investigate the function
$E_{0}(\gamma)$ for finite $N\gg 1$ and the zero BCs. Dimensional
reasoning implies that the energy $E_{0}$ of a 1D system of $N\gg 1$
point bosons  can be represented in the form \cite{ll1963}
\begin{eqnarray}
   E_{0}=Ng^{2}e_{0}(\gamma),
       \label{e0ll}\end{eqnarray}
where $e_{0}$ depends only on $\gamma$. We verified this property
for the zero BCs. To make it, we numerically found the numbers
$k_{j}$ for GS ($n_{j\leq N}=1$) from Eqs. (\ref{god-3}) at $N=10;
100; 1000$ and $c=0.01; 1; 100$. The analysis confirmed that, for a
fixed $N,$ the quantity $e_{0}$ in (\ref{e0ll}) depends only on
$\gamma$. If the system would transit from the gaseous (liquid)
state to a crystal one at a change in the parameters ($L$, $c$ or
$g$), then we would observe a jump on the curve $e_{0}(\gamma)$ (or
$\partial e_{0}(\gamma)/\partial \gamma$, or $\partial^{2}
e_{0}(\gamma)/\partial \gamma^{2}$) for some $\gamma=\gamma_{c}$.
Since the regimes $\gamma\ll 1$ and $\gamma \gg 1$ correspond to a
gas (Bogolyubov gas and the Tonks--Girardeau gas, respectively),
there should be two points $\gamma_{c},$ and they should be fairly
close to each other: $\gamma_{c}\sim 1$--$10$ for both ones.
However, the curves $e_{0}(\gamma)$ (see Fig. 5) contain no similar
peculiarities (we obtained these curves by changing $\gamma$ with a
sufficiently small step $\gamma\rightarrow 1.1\gamma$). The signs of
a smooth transition of the crossover type are also absent.

Thus, the ground state of the 1D system of spinless point bosons
corresponds to a liquid (gas) for any parameters of the system. In
the case of non-point interaction, the ground state of a 1D system
can correspond not only to liquid (gas) but also to a crystal (see
below).

In this work, we do not distinguish between gas and liquid. Indeed,
the main difference between a gas and a liquid is that liquid keeps
its volume (due to surface tension), while gas does not. This
difference disappears for the 1D system since it has no surface
(consequently, there is no surface tension). In principle, a gaseous
and liquid state could differ in the properties of the distribution
functions (e.g., the function $g_{2}(x_{1},x_{2})$). But then, the
transition from one phase to the other when $\gamma$ varies would
lead to a change of the function $e_{0}(\gamma)$. However, as we
noted above, features of such a change are absent. Therefore, we
believe that for the 1D system of point bosons, the gaseous and
liquid states do not differ.

\subsection{Mechanisms of formation of 1D Bose crystals}

As was mentioned in the Introduction, the crystalline regime was
found for the lowest state of a 1D Bose system with dipole-dipole
interaction
\cite{lozovik2005,reimann2010,zollner2011,zollner2011b,chatterjee2013,lode2018,chatterjee2019}.
In particular, for the singular potential
$U(x_{j},x_{l})=\frac{g_{d}}{(x_{j}-x_{l})^{3}}$ crystalline
solutions were obtained for the concentrations exceeding some
critical one: $n>n_{c}$ \cite{lozovik2005,zollner2011,zollner2011b}.
For the nonsingular potential
$U(x_{j},x_{l})=\frac{g_{d}}{(x_{j}-x_{l})^{3}+\alpha^{3}}$,
crystalline solutions were found for a strong interaction
($g_{d}>g_{c}$)
\cite{reimann2010,zollner2011,zollner2011b,chatterjee2013,lode2018,chatterjee2019}
and are almost independent of the magnitude of the trap field
\cite{chatterjee2019}. We remark that for $g_{d}>g_{c}$ there should
exist two solutions corresponding to the ideal crystal: the nodeless
solution obtained in
\cite{lozovik2005,reimann2010,zollner2011,zollner2011b,chatterjee2013,lode2018,chatterjee2019}
and a solution of the type (\ref{c-z}) with a large number of nodes.
We may expect that they are characterized by a similar structure of
$g_{2}(x_{1},x_{2})$, but by different energies.

The overall picture of the conditions under which the lowest state
of a 1D system of interacting bosons corresponds to a crystal is
still not completely clear. Crystalline solutions for a 1D system
were explored in works
\cite{gross1958,lozovik2005,mt2020,reimann2010,zollner2011,zollner2011b,chatterjee2013,lode2018,chatterjee2019}
(see also \cite{nep,kirz}). In this case, the analysis in
\cite{gross1958,nep,mt2020,kirz} is approximate. The numerical
solutions for few-boson systems
\cite{reimann2010,zollner2011,zollner2011b,chatterjee2013,lode2018,chatterjee2019}
and the exact solution in the present work are, seemingly, reliable.
Perhaps, a solution \cite{lozovik2005} by the Monte Carlo  method is
also close to exact. In other words, only the solutions for the
point-like and dipole-dipole
\cite{lozovik2005,reimann2010,zollner2011,zollner2011b,chatterjee2013,lode2018,chatterjee2019}
interactions are reliable. We conjecture that the main reason for
the crystallization of the lowest state of the dipole system is the
strong atomic repulsion on small (but nonzero) interatomic distances
$|x_{j}-x_{l}|$. Because of that repulsion, every two adjacent
particles strive to stay as far away from each other as possible,
which seemingly corresponds to a crystal state. In this case, the
crystalline ordering is energetically most preferred and arises in
the genuine nodeless GS. We suppose that the long-range character of
the interaction plays a small role in this instance: most of all,
crystallization will also occur in the case of a short (or mid)
range potential with radius $b\geq \bar{d}$ (e.g.,
$U(x_{j},x_{l})=\frac{g_{d}}{(x_{j}-x_{l})^{3}+\alpha^{3}}$ for
$|x_{j}-x_{l}|\leq b$, and $U(x_{j},x_{l})=0$ for $|x_{j}-x_{l}|>
b$); to our knowledge, only the case of $b=\infty$ was investigated
in the literature. In the case of point interaction, the above
mechanism does not work since a $j$-th and $l$-th particles interact
only when $|x_{j}-x_{l}|=0$. In that case, a crystal is created by
the nodes of WF. It is an entirely different mechanism. In addition,
the approximate analysis in \cite{gross1958,mt2020,kirz} indicates
that the crystalline solution can be energetically most favored
given the interaction is non-point: in this case, the Fourier
transform $\nu(k)$ of the potential is negative at some $k$'s, which
can make the crystal energetically preferred over a liquid. It
gives, in general, a third possible mechanism for crystal formation.
Under such a mechanism, the crystalline energy minimum in the space
of states may be absolute or local. If this minimum is absolute, the
third mechanism is equivalent (or close) to the first from the
mentioned above mechanisms. If the minimum is local,  the third
mechanism can be equivalent to the second (node) mechanism. To
clearly understand the conditions required for crystallization, one
needs to investigate the solutions for various potentials in detail.
It is a task for the future.

\section{Conclusion}

We note a few of the results obtained above. We have found the exact
crystalline solution for a 1D system of point bosons. It corresponds
to small and intermediate values of the coupling constant ($\gamma
\lsim 1$). The results of the earlier work \cite{syrwid2017}
indicate that such a solution should exist. It is of importance that
the ideal crystal corresponds to the wave function with nodes and
(at $N\gg 1$) is formed by the condensate of short-wave phonons.
Interestingly, at $0< \gamma < \infty$ the nodal structure of the
wave function is invariable at a change of $\gamma $ (we have
investigated $(N+1)^{N}$ lowest states for $N=2, 3$).

\section*{Acknowledgements}
The author acknowledges support by the National Academy of Sciences
of Ukraine grant ``Effects of external fields and spatial
inhomogeneities on the electronic properties of Dirac and
superconducting materials'' (program KPKVK 6541230).


       \end{document}